\documentclass[pra,twocolumn]{revtex4}
\usepackage{graphicx,amsmath,amssymb,float,bm,times}

\newcommand{\ek}{\epsilon_{\mathbf{k}}}

\newcommand{\Ek}{E_{\mathbf{k}}}

\newcommand{\createa}[1]{a^\dagger_{#1}}
\newcommand{\destroya}[1]{a^{\phantom \dagger}_{#1}}
\newcommand{\createb}[1]{b^\dagger_{#1}}
\newcommand{\destroyb}[1]{b^{\phantom\dagger}_{#1}}

\begin{document}

\title{The Nature of Superfluidity in Ultracold Fermi Gases Near
  Feshbach Resonances}

\author{Jelena Stajic$^1$, J. N. Milstein$^2$, Qijin Chen$^3$, M. L.
Chiofalo$^4$, M. J. Holland$^2$  and K.
  Levin$^1$}

\affiliation{$^1$ James Franck Institute and Department of Physics,
  University of Chicago, Chicago, Illinois 60637} 

\affiliation{$^2$ JILA, University of Colorado and National Institute
of Standards and Technology, Boulder, Colorado 80309}

\affiliation{$^3$ Department of Physics and Astronomy, Johns Hopkins 
University, Baltimore, Maryland 21218}

\affiliation{$^4$ Department of Mathematics and INFM, Group of Space
Mechanics, University of Pisa, Via Buonarroti 2 I-56127 Pisa, Italy}
\date{\today}

\begin{abstract}
  We study the superfluid state of atomic Fermi gases using
  a BCS-BEC crossover theory. Our approach emphasizes non-condensed
  fermion pairs which strongly hybridize with their (Feshbach-induced)
  molecular boson counterparts. These pairs lead to pseudogap effects above
  $T_c$ and non-BCS characteristics below. We discuss how these
  effects influence the experimental signatures of superfluidity.
\end{abstract}
\maketitle

There has been considerable interest in achieving ``resonance
superfluidity'' in an ultracold, extremely dilute, trapped Fermi gas
in which a Feshbach
resonance\cite{Holland,Timmermans} is used to tune the interatomic
attraction by variation of a magnetic field.  These systems allow
for the direct study of the crossover problem where the nature of the
superfluid transition changes from BCS to Bose-Einstein condensation (BEC).
Recently, several groups have observed superfluidity in the BEC 
regime\cite{Jin3,Grimm,Ketterle2}. This provides an exciting
opportunity for theoretical work which can address the entire 
range of behavior, in anticipation of future experiments.

In this paper we consider the case of a homogeneous gas as a first and
necessary step.
Our goals are to present a theory
for $T_c$, for the superfluid phase as well as
some signatures of the transition, which are shown to
be very different from conventional expectations based on BCS theory.
Using current understanding of bosonic superfluidity, it is useful
to begin with
a number of inferences about the nature of the superfluidity
induced when molecular bosons are admixed 
with fermions. For temperatures $T$ above $T_c$
sufficiently strong hybridization
($g b^\dagger_{{\bf q}} a^{\phantom\dagger}_{{\bf
q}-{\bf k}} a^{\phantom\dagger}_{{\bf
k}}$)
between molecular bosons ($b^\dagger_{\bf q}$)
and fermion pairs
will result in metastable ``pre-formed" pairs
($ a^\dagger_{{\bf
q}-{\bf k}} a^\dagger_{{\bf
k}}$). These 
are necessarily associated with a 
normal state excitation gap which represents the energy needed to
break them apart into fermionic quasi-particles  ($a^\dagger_{\bf k}$).
For $0< T \leq T_c$ the presence of non-condensed molecular bosons similarly
induces the
formation of non-condensed fermion pairs. These must be
present in addition to the usual
fermionic single particle excitations of the condensate. We may, thus,
quite generally infer that since the molecular bosons and the fermion pairs
are so strongly entangled, they must be treated on a similar footing.


Previous \cite{Milstein,Griffin} studies of the crossover problem in an
atomic Fermi gas at $T \ne 0$ have been based on the work \cite{NSR} of
Nozi\`eres and Schmitt-Rink (NSR). 
This approach, which effectively omits non-condensed
fermion pairs, addresses $T_c$ in a fashion which is not
manifestly compatible with the 
presumed ground state\cite{Leggett}. Moreover, pairing
correlations
are only included in the number equation and within an approximate
and evidently\cite{Serene,Balseiro} problematic fashion.
In this paper we address these shortcomings, but similarly base our
analysis of the standard BCS-like crossover ground state\cite{Leggett},
for the fermionic component.
We begin with the important observation that
the non-condensed fermion pairs are in chemical
equilibrium with \textit{both} the non-condensed molecular bosons and
the condensate. Consequently, they must satisfy
\begin{equation}
\mu_{pair}(T)= \mu_{boson}(T) = 0,\qquad T \leq T_c.
\label{eta_gg0}
\end{equation}
This last relation, which is a central equation of this paper, will
be rewritten shortly in a more concrete form.
To implement the physical picture discussed above, we follow the
theory in Ref.~\cite{Chen2}, extending it to include a Feshbach
resonance.  This approach was originally developed to treat high $T_c$
superconductors. The physics in the context of boson-fermion models is
shown here to be different from fermion-only models, although such
hybridization models have been also applied to the cuprates 
\cite{Micnas,Ranninger2}.

Our Hamiltonian\cite{Holland} is given by
\begin{eqnarray}
\label{hamiltonian}
H&-&\mu N=\sum_{{\bf k},\sigma}\epsilon_{\bf k}\createa{{\bf
    k},\sigma}\destroya{{\bf k},\sigma}+\sum_{\bf q}(\epsilon_{\bf
    q}^m+\nu)\createb{\bf q}\destroyb{\bf q}\nonumber\\ 
&+&\sum_{{\bf q},{\bf k},{\bf k'}}U({\bf k},{\bf k'})\createa{{\bf
    q}/2+{\bf k},\uparrow}\createa{{\bf q}/2-{\bf
    k},\downarrow}\destroya{{\bf q}/2-{\bf k'},\downarrow}\destroya{{\bf
    q}/2+{\bf k'},\uparrow}\nonumber\\ 
&+&\sum_{{\bf q},{\bf k}}\left(g({\bf k})\createb{{\bf q}}\destroya{{\bf
    q}/2-{\bf k},\downarrow}\destroya{{\bf q}/2+{\bf
    k},\uparrow}+h.c.\right). 
\end{eqnarray}
The sum in $\sigma$ runs over both spin states
$\{\uparrow,\downarrow\}$.  The free dispersion relations for fermions
and bosons are given by $\epsilon_{\bf k}=k^2/2m-\mu$ and $\epsilon_{\bf
  q}^m=E_{\bf q}^0-2\mu$ with $E_{\bf q}^0= q^2/2M$, respectively, where
we assume $\mu_\uparrow=\mu_\downarrow$, and $M=2m$ is the boson mass.
Here $\nu$ is the detuning of the resonance state, $U({\bf k},{\bf
  k'})=U\varphi_{\bf k} \varphi_{\bf k'}$ is the direct $s$-wave
interaction and $g({\bf k})=g\varphi_{\bf k}$ is the Feshbach coupling,
with the function $\varphi_{\bf k}^2 = \exp\{-(k/K_c)^2\}$ providing the
momentum cutoff.  Here we set $\hbar = k_B =1$.

The three propagators for fermion pairs, $t(Q)$, molecular bosons,
$D(Q)$, and single fermions, $G(Q)$ are coupled.
(Throughout we take the convention $\sum_K\equiv
T\sum_{\omega_n}\sum_{\bf k }$, where $K,Q$, etc.  are 4-vectors). 
With these definitions Eq.(\ref{eta_gg0}) can also be rewritten
\begin{equation}
D^{-1}(0) = t^{-1}(0) = 0, ~~ T \leq T_c.
\label{eta_gg01}
\end{equation}
The effective pairing interaction is given by \cite{Griffin} 
$\tilde{g}_{eff}(Q,K,K') =g_{eff}(Q)\varphi_{\bf k}\varphi _{\bf k'}$
with $g_{eff}(Q) \equiv U + g^2 D_0 (Q)$, where $D_0(Q) \equiv 1/ [
i\Omega_n- E_q^0 -\nu + 2 \mu ]$ is the non-interacting molecular boson
propagator with Matsubara frequency $\Omega_n$.  
We have from Eq.~(\ref{eta_gg01}) and the general form
of the T-matrix, the important result that
\begin{equation}
t^{-1} (0)= g_{eff}^{-1}(0)+\chi(0) = 0, \qquad T \le T_c  \;.
\label{eq:3}
\end{equation}
We now impose the reasonable constraints that at weak coupling
our results are compatible
with the $T$ dependence found in the BCS limit for $\Delta(T)$, and
for general coupling, the fermions are described by
the standard $T=0$ crossover state \cite{Leggett}. This, in turn,
constrains
the fermion pair susceptibility
$\chi(Q)$
\begin{equation}
\chi(Q) =  \mathop{\sum_K} G(K)G_0(Q-K)\varphi _{{\bf k}-{\bf q}/2} ^2, 
\label{chi}
\end{equation}
provided also the fermion self energy appearing in $G$ is
\begin{eqnarray}
 \Sigma(K) &=&  \mathop{\sum_Q} t(Q)G_0(Q-K)\varphi _{{\bf k}-{\bf q}/2} ^2.
\label{sigma_ggo} 
\end{eqnarray}
This form for $\chi$ and $\Sigma$ can also be derived by truncating the
equations of motion for Green's functions at the three-particle level.
This approach is also closely related to Hartree-approximated time
dependent Ginzburg-Landau theory\cite{JS}.

The T-matrix consists of two contributions: from the condensed ($sc$)
and non-condensed or ``pseudogap"-associated ($pg$) pairs. The
molecular bosons also contribute to the $T$-matrix via the effective
pairing interaction.
\begin{eqnarray}
t &=& t_{pg} + t_{sc} \label{t-matrix}\\
t_{pg}(Q)&=& \frac{g_{eff}(Q)}{1+g_{eff}(Q) \chi(Q)}, \qquad Q \neq 0 
\label{t-matrix_pg}\\
t_{sc}(Q)&=& -\frac{\tilde{\Delta}_{sc}^2}{T} \delta(Q) 
\label{t-matrix_sc}
\end{eqnarray}
where $\tilde{\Delta}_{sc}=\Delta _{sc}-g \phi _m$, with $\Delta
_{sc}=-U \sum _{\bf k}\langle a_{-{\bf k}\downarrow}a_{{\bf
    k}\uparrow}\rangle\varphi_{\bf k}$ and $\phi _m=\langle b_{{\bf
    q}=0} \rangle$. Without loss of generality, we choose 
order parameters $\tilde{\Delta}_{sc}$ and $\phi_m$ to be 
real and positive with $g<0$. The order parameter is a linear
combination of both paired atoms and condensed molecules.  These two
components are connected\cite{Kokkelmans}
by the relation $\phi _m=g \Delta _{sc}/[(\nu
-2\mu) U]$.  This implies that $\tilde{\Delta}_{sc}=-g_{eff}(0) \sum
_{\bf k}\langle a_{-{\bf k}\downarrow}a_{{\bf k} \uparrow} \rangle
\varphi_{\bf k}$, as expected. Following Eq.~(\ref{eta_gg01}), for $ T
\le T_c$ we may approximate\cite{Maly1} Eq.~(\ref{sigma_ggo}) to yield
a BCS-like self-energy with $\Delta^2 \equiv \tilde{\Delta}_{sc}^2 +
\Delta_{pg}^2$
\begin{equation}
\Sigma (K)\approx -G_0 (-K) \Delta^2 \varphi _{\bf k}^2
\label{self_energy_gg0}
\end{equation}
where we define the pseudogap $\Delta_{pg}$
\begin{equation}
\Delta_{pg}^2 \equiv -\sum_{Q\ne 0} t_{pg}(Q) \;.
\label{delta_pg}
\end{equation}
At $T=0$, $\Delta_{pg} =0$, so that $\tilde{\Delta}_{sc}(0) = \Delta(0)$.
Equation (\ref{eq:3}) can be written
as
\begin{equation}                
g_{eff}^{-1}(0) +  \mathop{\sum_{\bf k}} \frac{1 - 2 f(E_{\bf k})}{2
  E_{\bf k}} \varphi _{\bf k}^2 = 0, \qquad  T \le T_c\;,
\label{eq:gap_equation}
\end{equation}
where $\Ek = \sqrt{ \ek ^2 + \Delta^2 \varphi _{\bf k}^2}$. Equation
(\ref{eq:gap_equation}) has the form of the conventional BCS equation,
\textit{but the full excitation gap $\Delta$, as distinguished from the
  order parameter, $\tilde{\Delta}_{sc}$, appears in the dispersion
  $E_{\bf k}$}.

We show next that this equation for the excitation gap $\Delta(T)$
coincides with the equivalent condition on the molecular bosons,
given in Eq.(\ref{eta_gg01}). 
Taking the same pair susceptibility $\chi$ in the boson self-energy
$\Sigma_B$, we obtain
\begin{equation}
 \Sigma_B (Q) \equiv -g^2 \chi(Q) / [ 1 + U \chi(Q) ]
\end{equation}
so that the boson propagator is
\begin{equation}
D(Q) \equiv  \frac{1}{i\Omega_n - E_q^0 - \nu + 2 \mu  - \Sigma_B(Q)}.
\end{equation}
After some algebra, Eq.(\ref{eta_gg01})
leads directly to Eq.~(\ref{eq:gap_equation}).
We may now calculate the number equation from the propagators involved.
The number of non-condensed molecular bosons is given directly by $n_b =
-\sum_Q D(Q)$.  For $ T \le T_c$, the number of fermions is
\begin{equation}
n_f = \sum _{\bf k} \left[ 1 -\frac{\ek}{\Ek} +2 \frac{\ek}{\Ek}
  f(\Ek)\right] \;,
\label{eq:number_equation}
\end{equation}
as follows from the condition $n_f = 2 \sum_K G(K)$ with a BCS-like self
energy $\Sigma(K)$.  The \textit{total} number (n) of fermions is then
\begin{equation}
n_f + 2 n_b + 2 n^0_b = n \;,
\label{number_equation}
\end{equation}
where $n^0_b=\phi _m^2$ is the number of molecular bosons in the
condensate.

We now have a closed set of equations for our resonance system which
require numerical solution.  The cutoff function $\varphi_{\bf k}$ introduces a
renormalization of $U$, $g$, and $\nu$.  Extending the derivation given
in Ref.~\cite{Kokkelmans} to a nonconstant, separable potential,
we find that $U=\Gamma U_0$, $g=\Gamma g_0$, and $\nu=\nu_0+\alpha\Gamma
g_0^2$.  Here $U_0=4\pi a_{bg}/m <0$ where $a_{bg}$ is the
background scattering length and $g_0$ is the physical scattering parameter 
reflecting the width of the Feshbach resonance.
The scaling factor $\Gamma=1/(1-\alpha U_0)$ and $\alpha=m
K_c/(4\pi^{3/2})$.

At the physical level, the essential distinction between this and
previous crossover studies based on the NSR approach \cite{Griffin2}
below $T_c$ is associated with non-condensed pair excitations
\cite{Chen3} of the superfluid.  At a more formal level, it should be
noted that there is an important difference between Eq.~(\ref{chi}) and
previous related work \cite{NSR}.  This NSR approach presumes that there
are two bare Green's functions $G_0$ in $\chi(Q)$: the particles acquire
a self energy from the pairs, but these self energy effects are not fed
back into the propagator for the pairs.

To evaluate $T_c$, $\Delta (T)$, $\mu (T)$ as well as
$\tilde{\Delta}_{sc}(T)$ and other transport properties \cite{Chen2,JS}
below $T_c$, we solve Eqs.~(\ref{delta_pg}), (\ref{eq:gap_equation}) and
(\ref{number_equation}) for fixed $U_0$, $g_0$, and $\nu_0$ and a
sufficiently large cutoff $K_c$.  Due to the divergence of the T-matrix
[Eq.~(\ref{eq:3})], we may Taylor expand the quantity $\chi(Q)$ in
Eqs.~(\ref{number_equation}) and (\ref{delta_pg}) to first order in
$\Omega$ and $q^2$: $\chi (Q)=\chi(0)+a_0 (i\Omega _n-B_0 q^2)$
\cite{JS}, and expand $g_{eff}(Q)$ and $\Sigma_B(Q)$ similarly. This
considerably simplifies the analysis.

\begin{figure}
\includegraphics[angle=0,width=3.2in]{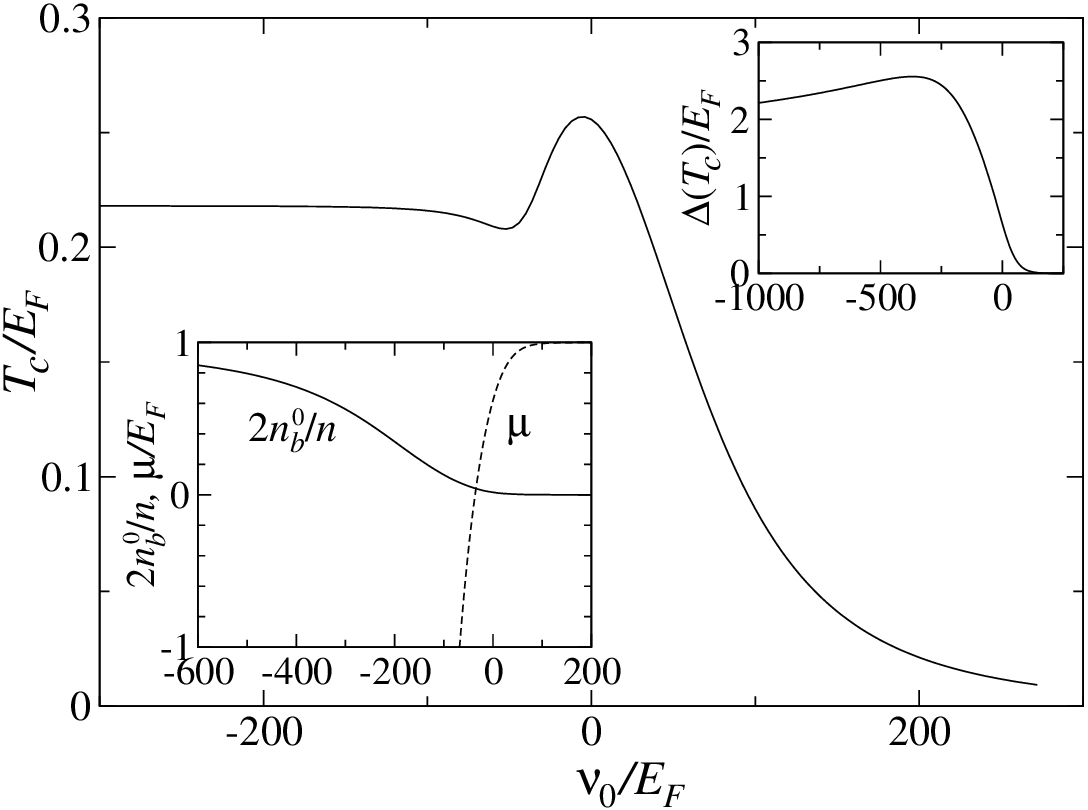}
\caption{$T_c$ vs detuning $\nu_0$. Upper inset
  plots the pseudogap at $T_c$ vs $\nu_0$, and lower inset
  plots molecular boson contribution to the condensate weight and
  fermionic chemical potential at $T=0$.}
\label{fig:tc}
\end{figure}    

We begin with calculations of $T_c$ as a function of $\nu_0$ which are
plotted in Fig.~\ref{fig:tc}.  Here we take $g_0=-42 E_F/k_F^{\frac{3}{2}}$ and
$U_0= -3 E_F/k_F^3$ as is expected to be reasonably indicative of the
behavior for $T_c$ in currently trapped atomic Fermi gases.  
For $\nu_0 \rightarrow - \infty$ only molecular bosons are present and
$T_c$ approaches the ideal BEC limit. As $ \nu_0 \rightarrow + \infty $,
molecular bosons become irrelevant and the asymptote of the curve is
dictated by the behavior of fermions in the presence of $U$.  In this
paper we have chosen $U_0$ deliberately to be small so that the $\nu_0
\rightarrow \infty$ limit is close to BCS.  

Indicated in the upper right inset is the behavior of $\Delta(T_c)$ as a
function of $\nu_0$. The lower left inset shows a plot of the molecular
Bose condensate weight $n_b^0$, and the fermionic chemical potential
$\mu$ as a function of $\nu_0$.  For positive, but decreasing $\nu_0$,
$T_c$ follows the BCS curve until the ``pseudogap" or $\Delta(T_c)$
becomes appreciable. At this point the molecular Bose condensate also
becomes appreciable. After its maximum (at slightly negative $\nu_0$),
$T_c$ decreases, as does $\mu$, until $\mu = 0$.  Beyond this point,
towards negative $\nu_0$, the system is effectively bosonic. The
condensate consists of two contributions, although the weight of the
fermion pair component rapidly disappears, as can be inferred
from the lower inset.
Similarly $T_c$ rises, although slowly, towards the ideal BEC asymptote,
following the inverse effective boson mass.
%
%
The corresponding curve based on the NSR approach has only one extremum,
but nevertheless the overall magnitudes are not so
different\cite{Griffin,Milstein}.  There are, however, key differences
between the behavior of the fermionic
excitation gap $\Delta(T_c)$ and its high
$T_c$ counterpart.  Because all of the condensate comes from molecular
bosons in the strict BEC limit, $\Delta(T_c)$ reaches a maximum (where
the molecular bosonic and fermionic weights are comparable), and then
decreases, as $\nu_0$ decreases.  This is in contrast to fermion-only
based models\cite{Chen2}
where this parameter increases indefinitely as the attractive
coupling becomes stronger.

\begin{figure}
\includegraphics[angle=0,width=3.2in]{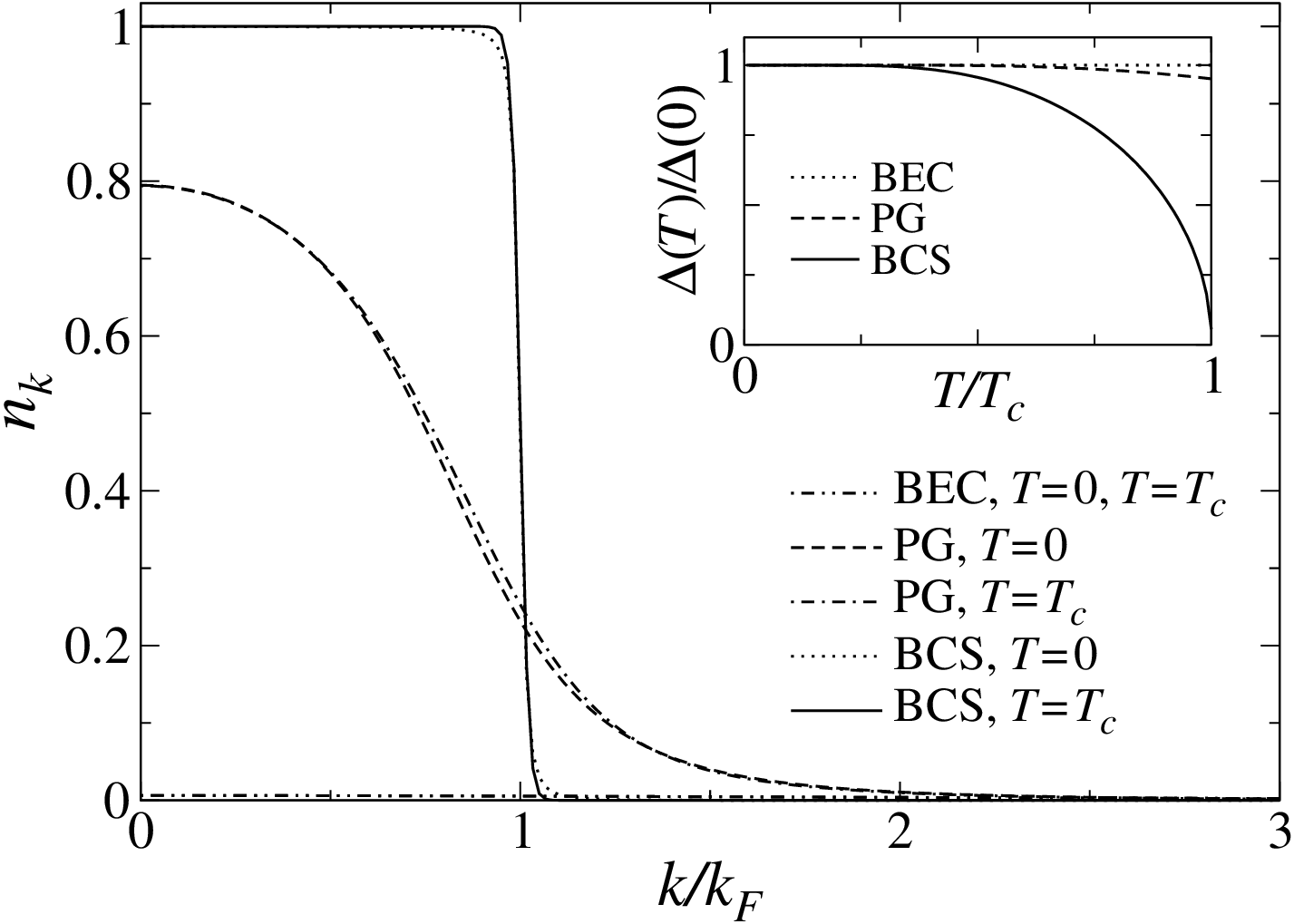}
\caption{Fermionic momentum distribution function at $T=0$ and $T=T_c$
for the three regimes.  $PG$ corresponds to maximal $T_c$.
Inset plots the $T$ dependent excitation gap below
$T_c$.}
\label{fig:nk}
\end{figure}
     
In the inset to Fig.~\ref{fig:nk}, we plot the temperature dependence of
the normalized excitation gap $\Delta(T)/\Delta(0)$ for three values of
$\nu_0/E_F = -200, -5, +200$.  The second value corresponds roughly to the
maximum in $T_c$, where 
pseudogap effects are apparent.  The first and last are illustrative of
the BEC and BCS limits respectively.  We, thus, refer below to these
three values as $BEC$, $PG$ and $BCS$ cases. While not shown here, $\mu$
is positive for the latter two and negative for the first.  The order
parameter, $\tilde{\Delta}_{sc}$, is not plotted here, but for all three
cases it is rather close to the solid line in the inset.  That
$\Delta(T)$ is relatively constant with $T$ through the superfluid
transition is to be expected in the presence of ``pre-formed pairs", as
for $\nu_0/E_F = -5$ and $-200$.

It should be stressed that $T_c$ is only apparent in $\Delta(T)$ for the
BCS case.  To underline this point, in the main body of
Fig.~\ref{fig:nk} we plot the fermionic momentum distribution function
$n_k$ \cite{Stringari}, which is the summand in
Eq.~(\ref{eq:number_equation}), at $T=0$ and $T=T_c$.  The fact that
there is very little change from $ T = 0$ to $ T= T_c$ makes the
important point that this momentum distribution function is not a good
indicator of phase coherent pairing.  For the $PG$ case, this, in turn,
derives from the fact that $\Delta(T)$ is nearly constant.  For the BEC
limit the excitation gap, which is dominated by $\mu$, similarly, does
not vary through $T_c$. In the BCS regime, $\Delta(T)$ is sufficiently
small as to be barely perceptible on the scale of the figure.  In order
to address the measurable particle density distribution, these
observations will have to be incorporated into previous BCS-based (i.e.,
$\Delta \equiv \tilde{\Delta}_{sc}$) calculations \cite{Hulet,Chiofalo},
albeit generalized to include inhomogeneity effects.

This pseudogap-based phenomenology is well documented in the high $T_c$
superconductors, although for these materials, penetration depth data
(with no analogue here) are direct probes of the transition to
superconductivity.  Interestingly, densities of state measurements in
the cuprates also show some indications of when order is well
established.  To see how phase coherence enters in the
atomic physics context, we relax the approximation in
Eq.~(\ref{self_energy_gg0}) by noting that incoherent or finite momentum
pairs ($pg$) are distinguishable from coherent or zero momentum pairs
($sc$) through lifetime effects in the self energy. For the $pg$ term we
write $\Sigma_{pg}(\omega, {\bf k}) \approx \Delta_{pg}^2 / ( \omega +
\epsilon_{\bf k} + i \gamma )$.  For simplicity, we treat $\gamma$ as a
phenomenological parameter which is independent of $T$.  This
distinction between $pg$ and $sc$ is required to arrive at general
thermodynamic signatures (not discussed here) of the transition.

\begin{figure}
\includegraphics[angle=0,width=3.2in]{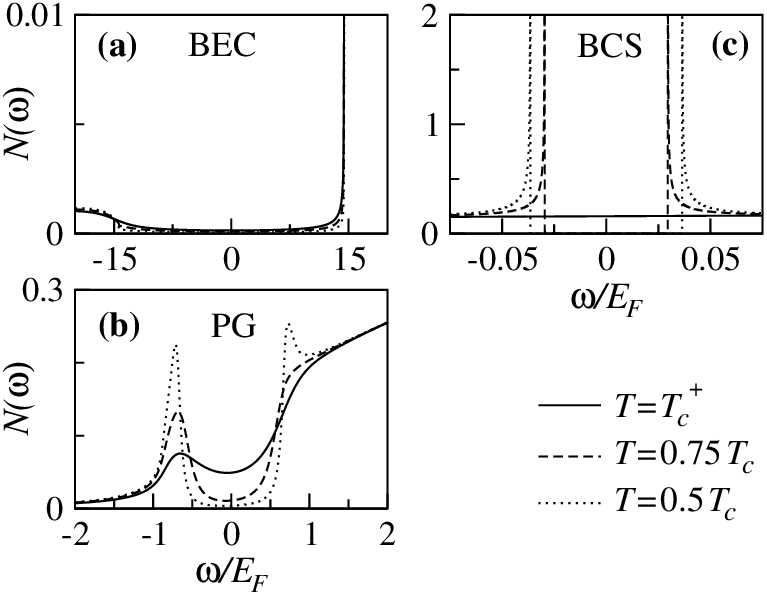}
\caption{Fermionic density of states vs energy for the three regimes
  at three indicated temperatures. Note the difference in the scales.}
\label{fig:dos}
\end{figure}  

Figures \ref{fig:dos}(a)-\ref{fig:dos}(c) show the resulting density of
states $N(\omega)=-2 \sum_{\bf k} {\rm Im}\, G(\omega+i0,{\bf k})$ 
and correspond to the $BEC$, $PG$, and $BCS$ cases, as in
Fig.~\ref{fig:nk}. The temperatures shown are just above $T_c$, and for
$ T = 0.75 T_c$ and $ T = 0.5 T_c$.  The $BCS$ case indicates an abrupt
transition at $T_c$, but with a very small gap and very low $T_c$, which
may be hard to observe experimentally.  The $BEC$ case shows very little
temperature dependence throughout, since the fermions are well gapped at
all temperatures.  Only the $PG$ case, where $T_c$ is maximal, indicates
the presence of superfluidity, not so much at $T_c$ but once superfluid
order is well established at $ T = 0.5 T_c$, through the presence of
sharper coherence features, much as seen in the cuprates.

These plots have important implications for interpreting predicted
signatures of superfluidity, such as those \cite{Torma} based on laser
probing of ``atomic Cooper pairs", where it has been argued that there
is a conceptual analogy between normal metal-superconductor tunneling
[which measures $N(\omega)$], and resonant scattering of laser light.
The present work introduces a caution in interpreting future atomic trap
experiments: \textit{Because of the presence of a pseudogap, the
  signatures of superfluid onset are not as simple as in BCS or the
  related Bogoliubov-de Gennes theory. In general one has to distinguish
  between the superfluid order parameter and the fermionic excitation
  gap}.  Nevertheless, superfluid coherence appears to be visible as
fine structure effects in the fermionic density of states.  While the
fermionic contributions [via $N(\omega)$, $n_k$] do not provide a clear
indication of superfluidity, they do establish the nature of the pairing
regime: BCS, BEC or PG --- a ``pseudogapped superfluid".

In addition to investigating and revisiting experimental signatures, the
main theoretical contributions of this paper are to establish the
presence and role of pre-formed fermion pairs at and above $T_c$, which
evolve into non-condensed pairs below $T_c$.  These 
pairs hybridize strongly with their molecular boson counterparts
associated with the Feshbach resonance. In this context we present
a generalized mean field treatment of the broken symmetry phase for
ultracold fermionic atoms, which unlike other $T \neq 0$ approaches,
connects smoothly to the conventional\cite{Leggett} crossover ground
state.
Because the Feshbach resonance has no natural analogue in high $T_c$
systems, it remains to be seen whether the differences in these two
systems will outweigh the similarities.

We acknowledge useful discussions with D. Jin and R. Hulet.  This work
was supported by NSF-MRSEC Grant No.~DMR-0213745 (JS and KL), NSF Grant
No.~DMR0094981 and JHU-TIPAC (QC), by the U.S. Department of Energy
(JNM), by SNS Grant PRIN 2002, Pisa, Italy
(MLC), and by the National Science Foundation (MLC,MJH).


\end{document}